\input harvmac

\Title{}{A Simple Solution to the Strong CP Problem}

\centerline{\bf Sheldon Lee Glashow}\bigskip

\centerline{Department of Physics}
\centerline{Boston University}
\centerline{Boston, MA 02215}
\vskip .4in

We propose a minimal modification  of the standard model,
remarkable in its simplicity,  which may
solve the strong CP problem. It employs  three Higgs  doublets 
 with interactions taken to be  invariant under a flavor symmetry.
Both CP and the flavor symmetry are softly broken by Higgs boson
mass terms. In tree approximation,  quark mass matrices are triangular and
 $\arg\det M$  vanishes. Radiative corrections lead to a tiny and
tolerable value of $\overline\theta$.

\Date{10/01}

The problem is simply stated.  The minimal three-family standard
model involves two distinct  CP-violating phases: the $\delta$ parameter
in the Kobayashi-Maskawa matrix and $\overline \theta$, the measure of strong
CP violation, both of which can be generated by complex terms in the quark
mass matrices.  The former angle is of order unity, whereas the latter is
known to be less than $10^{-9}$. The 
strong CP problem arises if this disparity is
regarded as other than coincidental.

\medskip

There have been many proposed resolutions to this problem of which three
classics are: a massless up quark~\ref\rac{For a recent analysis, see: A.G.
  Cohen, D.B. Kaplan and A.E. Nelson, JHEP 9911 (1999) 27.}, an invisibilized
version of the Peccei-Quinn axion~\ref\rpq{R.D. Peccei and H.R. Quinn, Phys.
  Rev. Lett. 38 (1977) 387.}, and the Barr-Nelson implementation of
spontaneous CP violation~\ref\rbn{S.M. Barr, Phys. Rev. Lett. 53 (1984)
  329\semi A.E.  Nelson, Phys. Lett. B136 (1984) 387.}. Although none of
these $\overline \theta$--avoidance mechanisms are decisively excluded,
neither is convincingly true. Consequently there have been many other
suggested remedies which would be too tedious to enumerate.  Some models
appeal to supersymmetry~\ref\rms{{\it E.g.,} G. Hiller and M. Schmaltz, {\tt
    hep-ph/0105254.}}, others appeal to technicolor~\ref\rkl{{\it E.g.,} K.D.
  Lane, {\tt hep-ph/0106328.}}, and still others invoke extra
dimensions~\ref\rxd{{\it E.g.,} M. Chaichan and A.B. Kobakhidze, {\tt
    hep-ph/0011376.}}.  In addition, many solutions have been put forward in
which CP is softly broken~\ref\rsg{{\it E.g.,} P. H. Frampton and S.L.
  Glashow, Phys. Rev. Lett., 87 (2001) 011801\semi H. Georgi and S.L Glashow,
  Phys. Lett., B451 (1999) 372.}. The last-cited models (of which I am a
coauthor) are decisively ruled out by currently available experimental data.
In this note I present  yet  another model making use of soft CP violation, one
which is both simpler than its predecessors and seemingly compatible with
experiment.  \medskip

We begin by assigning a flavor quantum number $F$ to each  quark
family. We  assign $F=+1$ to the right- and left-handed
quarks of a nominal first family, $F=0$ to those of a second family, and
$F=-1$ to those of the third family. (Similar $F$ assignments could be made
to the leptons.)  In addition, we introduce three doublets of Higgs bosons,
$h_0$, $h_1$ and $h_2$, where the subscripts indicate their $F$ quantum
numbers. Each of these doublets contributes both to the up and down quark
mass matrices. Although the Glashow-Weinberg 
constraint~\ref\rgw{S.L. Glashow and S. Weinberg, Phys. Rev. D15 (1977) 1958.}
is not respected, unacceptable  flavor-changing effects 
should be  avoidable with a
modest degree of fine tuning.

\medskip All dimension-4 terms in the Lagrangian are assumed to be both CP
and flavor invariant. As a result of this hypothesis, the Yukawa couplings of
the $h_i$ to quarks (as well as their quartic self-couplings) must be real
and must conserve the flavor quantum number $F$. Furthermore, the
CP-violating Chern-Simons term cannot be present in  the Lagrangian. 
Quadratic expressions in the Higgs fields are not constrained by the above
hypothesis. These dimension-2 mass terms are permitted to be complex and
$F$-violating. This is our proposed mechanism for the explicit but soft
violation of both CP invariance and the flavor symmetry.\medskip

  The parameters of
the Higgs portion of the Lagrangian are chosen so  that each of the three
electrically neutral Higgs bosons develops a complex vacuum expectation value
(vev). An appropriate weak-hypercharge rotation lets us choose
 $\langle h_0\rangle$ to be real with no loss of generality.
The sum of the squared magnitudes of the vevs
is constrained to take its conventional
electroweak value.  We assume that all  three vevs are similar in magnitude to
avoid the appearance of light surviving bosons ({\it i.e.,}
with masses  less than $\sim
100$~GeV) such as are known not to exist. Observed violations of 
CP invariance are caused by  the complexity of the Higgs  vevs, which lead
to the irremoveable
 complexity of the quark mass matrices $M_U$ and
$M_D$. Furthermore, the $F$ invariance of the Yukawa couplings ensures
that in tree approximation $M_D$ is an upper triangular matrix whilst
$M_U$ is lower triangular. All 
 diagonal entries of these matrices are real
 and the phases of the off-diagonal  entries are  constrained in
 a manner to be discussed. 
\medskip

The triangular mass matrices we obtain are sufficiently general to produce
any spectrum of quark masses and can result in any desired Kobayashi-Maskawa 
matrix. Furthermore, the determinants of these mass matrices are evidently
real, so that there is no strong CP problem in tree approximation.

\medskip

With one possible identification of the observed quarks with the
flavor quantum number, we obtain:
\eqn\eud{M_D=\pmatrix{m_d&\epsilon_{12}&\epsilon_{13}\cr
           0& m_s& \epsilon_{23}\cr 
           0&0& m_b\cr} \quad\quad{\rm and}\quad\quad
          M_U=\pmatrix{ m_u&0&0\cr
            \epsilon_{21}^*&  m_c&0\cr
        \epsilon_{31}^*&\epsilon_{32}^*& m_t\cr}\,,}
where the $ m_i$ are the quark masses in the absence of  mixing. 
Quark mixing is induced by smaller  off-diagonal entries, whose complex
phases are constrained as follows:
\eqn\ephases{
\arg\epsilon_{12}=\arg\epsilon_{21}=\arg\epsilon_{23}=
\arg\epsilon_{32}\,,\quad\ {\rm and}\quad\ 
\arg\epsilon_{13}=\arg\epsilon_{31}\,.
 }
Note that the complexity of the  mass matrices is  irremoveable except
under the coincidental circumstance that 
\eqn\echi{\chi\equiv \arg(\epsilon_{12}^2\,\epsilon_{13}^*)=0\,.}  

\medskip

It is illustrative to estimate the Kobayashi-Maskawa parameters in the
small-mixing approximation. We obtain:
\eqn\ekm{\theta_{12}\approx
  \vert\epsilon_{12}/m_s-\epsilon^*_{21}m_u/m_c^2\vert\,,\quad
  \theta_{23}\approx \vert\epsilon_{23}/m_b-\epsilon^*_{32}m_c/m_t^2\vert,\,
\quad \theta_{31}\approx
  \vert\epsilon_{13}/m_b-\epsilon^*_{31}m_u/m_t^2\vert\,,} 
with the $\delta$ parameter assuming any desired value. Note that the KM
parameters are  essentially  determined by the off-diagonal terms of $M_D$,
provided that the $\epsilon_{ij}$ with $j>i$ are comparable to those with
$i>j$. Putting in numbers for quark masses and mixings, we obtain the
rough estimates:
\eqn\est{\epsilon_{12}\approx 25{\rm ~ MeV}\,,\quad\ 
\epsilon_{13}\approx 13{\rm  ~MeV}\,,\quad\ \epsilon_{23}
\approx 150{\rm ~MeV}\,.}

\medskip

While our model does not suffer a strong CP problem in tree approximation,
radiative  corrections  to quark masses will modify  the  quark
 mass matrices and  can  thereby lead to a non-vanishing value for
$\overline\theta$. Are these corrections   small enough to avert a problem?
An examination of one-loop corrections to the quark mass matrices reveals
that the only such terms are those  explicitly involving  $\chi$, as defined
in Eq.~\echi. The most threatening term by far is a
complex  contribution to the up-quark
mass mediated by charged Higgs bosons, which may be  estimated to yield:
\eqn\ethetabar{\Delta \overline\theta=
\left({1\over 4\pi}\right)^2\, {\epsilon_{13}\,\epsilon_{23}^*\,\epsilon_{21}^*
\over ({\rm vev})^2 \,m_u}\,K\,,}
where $K$ is a dimensionless integral which could be small for an appropriate
choice of the spectrum and mixing pattern of the Higgs bosons. In any case,
its prefactor is of order $10^{-9}$.
\

\medskip

Several distinctive features of this model may be testable. In particular, it
requires the existence of two surviving 
charged Higgs bosons, which should be readily
detectable at a future linear collider. Furthermore, should one or both of
these particles lie much below the
top quark mass,  we should expect a significant (although not readily
calculable)  branching
ratio for the decay $t\rightarrow b+h^+$. The predicted existence of five
neutral bosons should offer interesting challenges to experimenters. Although
flavor violation via their exchange can be made small, it could be large enough
to yield  measureable departures from the standard-model
description of CP violation. 

\bigskip
This work was supported in part by the National Science Foundation,
under grant NSF-PHY-0099529.

\listrefs
\bye